\documentclass[3p,times,twocolumn]{elsarticle}

\usepackage{ecrc}

\volume{00}

\firstpage{1}

\journalname{Subatomic Particles and Cosmology}

\jid{nppp}

\jnltitlelogo{Subatomic Particles and Cosmology}

\usepackage{amssymb}

\usepackage{lineno}

\usepackage{booktabs}
\usepackage{amsmath}
\usepackage{tabularx}
\usepackage{subcaption}

\usepackage[figuresright]{rotating}

\begin{document}

\begin{frontmatter}



\dochead{}

\title{Charmed baryon decays at Belle and Belle II}
\author{Jaeyoung Kim$^{a}$ on behalf of the Belle II Collaboration}
\address[a]{Department of Physics, Yonsei University, Seoul 03722, Republic of Korea}

\begin{abstract}
Belle and Belle II experiments have collected $e^+e^-$ collision data with center-of-mass energies at or near the $\Upsilon(4S)$ resonance. 
Using total $1.4\,\mathrm{ab}^{-1}$ combined dataset, we present new measurements of branching fractions for $\Xi_c^{0/+}$ and $\Lambda_c^+$ baryons, including several first observations. 
Additionally, we report the initial search for $CP$ violation in singly Cabibbo-suppressed three-body decays of charmed baryons, providing a test of $U$-spin symmetry.
\end{abstract}


\end{frontmatter}


\section{Introduction}
Charmed baryons provide a unique system for studying the dynamics of the interplay between strong and weak interactions.
Operating at the KEKB~\cite{Kurokawa:2001nw,Abe:2013kxa} and SuperKEKB~\cite{Akai:2018mbz} $e^+e^-$ colliders, Belle~\cite{Belle:2000cnh} and Belle II~\cite{Belle-II:2010dht} have collected a total of $1.4 \text{ ab}^{-1}$.
These experiments generally reconstruct charm baryons produced in the continuum $e^+e^- \to c \bar{c}$ process.
The process offers a relatively clean experimental signature for the reconstruction of charm baryon particles.

These proceedings summarize recent results of charm baryon decays using the datasets from the Belle and Belle II (Run 1).
Specifically, we report new measurements of branching fractions and searches for rare decays in various $\Xi_c^{0/+}$ and $\Lambda_c^+$ decays.
Furthermore, we present first searches for $CP$ violation in singly Cabibbo-suppressed three-body decays of $\Xi^+_c$ and $\Lambda_c^+$ baryons.
 
\section{Branching Fractions and Rare Decays}
\subsection{$\Xi_c^0 \to \Xi^0h\;(h=\pi^0,\eta,\eta^{\prime})$}
The two-body decays $\Xi_c^0 \to \Xi^0 h$(where $h=\pi^0,\eta,\eta^{\prime}$), followed by $\Xi^0\to \Lambda(\to p\pi^-) \pi^0$, are Cabibbo-favored (CF) processes to which only nonfactorizable amplitudes contribute.
These amplitudes arise from internal $W$-emission and $W$-exchange quark-level processes, making these modes sensitive tests for various theoretical approaches, such as pole models~\cite{Cheng:1993gf, Zou:2019kzq} and $\mathrm{SU}(3)_F$ flavor symmetry~\cite{Geng:2019xbo}.

\begin{figure}[t]
    \centering
    \includegraphics[width=1\columnwidth]{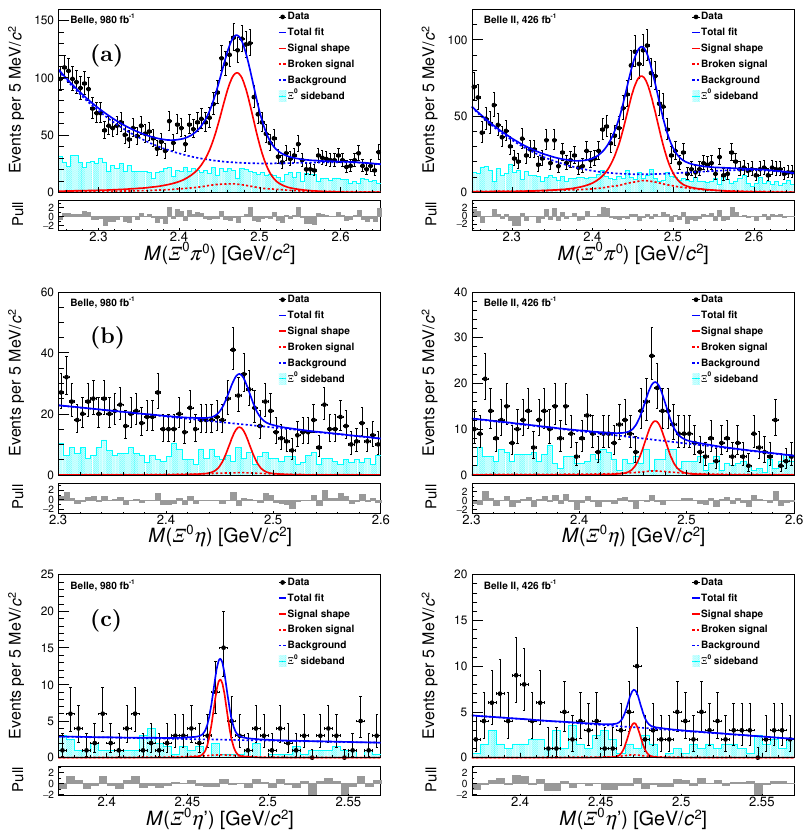}
    \caption{Invariant mass distributions for the reconstructed (a)~$\Xi_c^0 \to \Xi^0\pi^0$, (b)~$\Xi^0\eta$, and (c)~$\Xi^0\eta^{\prime}$ candidates. The blue solid curves represent the total fit, while the dashed lines indicate the background components. The left and right columns correspond to Belle and Belle II samples, respectively. Figure adapted from Ref.~\cite{Belle:2024ikp}.}
    \label{fig:xic_fits_cf}
\end{figure}

This analysis~\cite{Belle:2024ikp} utilizes the $980 \text{ fb}^{-1}$ data collected by the Belle and a $426 \text{ fb}^{-1}$ data from the Belle II.
The signal yields are extracted from unbinned extended maximum-likelihood fits to the invariant mass spectra of reconstructed $\Xi_c^0$ candidates as shown in Fig.~\ref{fig:xic_fits_cf}.
To determine absolute branching fractions, the decays are normalized to the well-measured $\Xi_c^0 \to \Xi^- \pi^+$ mode.

The resulting measurements, summarized in Table~\ref{tab:xic_branching_ratios}, represent the first observations of these three decay modes. The results are found to be consistent with theoretical predictions based on $\mathrm{SU}(3)_F$ flavor symmetry-breaking model~\cite{Zhong:2022exp}, whereas our measurements marginally disagree with certain predictions provided by the covariant confined quark model~\cite{Korner:1992wi,Ivanov:1997ra}.

While the magnitude of systematic effects is decay-dependent, the analysis is generally dominated by the modeling of broken-signal shapes and the reconstruction efficiencies of $\pi^0$ and $\gamma$.
Specifically, the $\Xi^0\pi^0$ mode is most sensitive to $\pi^0$ efficiency.
In contrast, the $\Xi^{0}\eta$ channel is primarily limited by the parameterization of broken-signal candidates and background shapes, whereas $\Xi_0 \eta^{\prime}$ is driven by both $\pi^0$ and $\gamma$ reconstruction and background modeling.

\begin{table}[h!tbp]
\centering
\caption{Measured branching fraction ratios for $\Xi_c^0$ decay modes.}
\begin{tabular}{lccc}
\toprule
Mode & Branching Ratio \\
\midrule
$\mathcal{B}(\Xi_c^0 \to \Xi^0\pi^0)/\mathcal{B}(\Xi_c^0 \to \Xi^-\pi^+)$ & $0.48 \pm 0.02 \pm 0.03$ \\
$\mathcal{B}(\Xi_c^0 \to \Xi^0\eta)/\mathcal{B}(\Xi_c^0 \to \Xi^-\pi^+)$  & $0.11 \pm 0.01 \pm 0.01$ \\
$\mathcal{B}(\Xi_c^0 \to \Xi^0\eta^{\prime})/\mathcal{B}(\Xi_c^0 \to \Xi^-\pi^+)$ & $0.08 \pm 0.02 \pm 0.01$ \\
\bottomrule
\end{tabular}
\label{tab:xic_branching_ratios}
\end{table}

\begin{figure}[t]
    \centering
    \includegraphics[width=1\columnwidth]{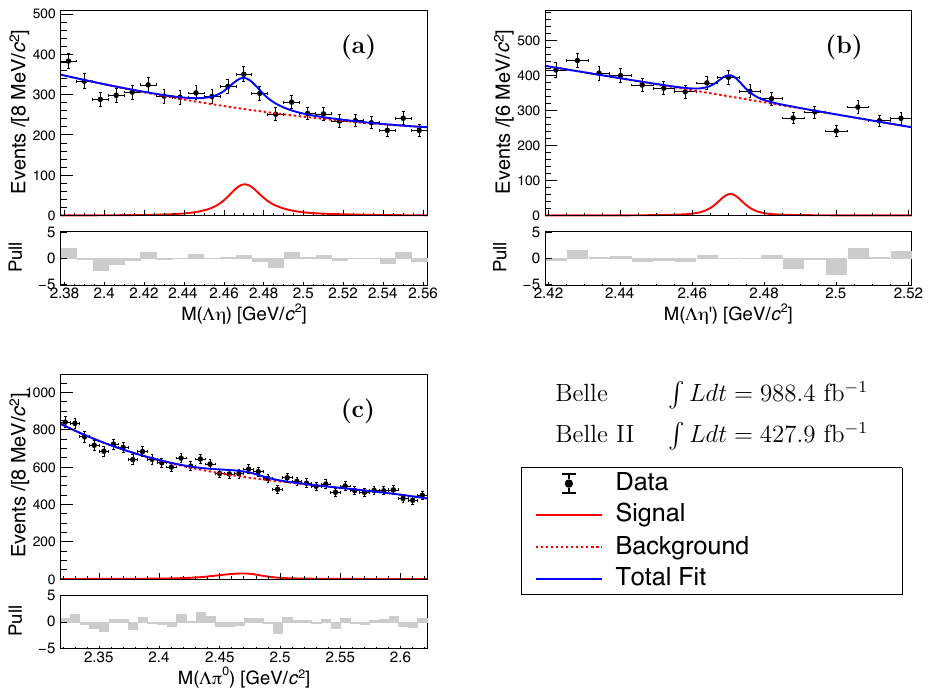}
    \caption{Invariant mass distributions for (a) $\Xi_c^0 \to \Lambda \eta$, (b) $\Lambda \eta^{\prime}$, and (c) $\Lambda \pi^0$ candidates reconstructed from the combined Belle and Belle II data samples. The total fit results are denoted by solid blue curves, with the background contributions represented by red dashed lines. Figure adapted from Ref.~\cite{Belle:2025qrq}.}
    \label{fig:xic_fits_scs}
\end{figure}

\subsection{$\Xi_c^0 \to \Lambda h\;(h=\pi^0,\eta,\eta^{\prime})$}
The preliminary results presented at this conference have since been published in~\cite{Belle:2025qrq}.
These decays are classified as singly Cabibbo-suppressed (SCS) processes.
However, unlike the $\Xi_c^0\to \Xi^0 h$ decays discussed in the previous section, $\Xi_c^0\to\Lambda h$ modes can proceed via both factorizable (internal/inner $W$-emission) and nonfactorizable ($W$-exchange) amplitudes.
The experimental measurement of these branching fractions is essential to test the validity of different theoretical frameworks, including $\mathrm{SU}(3)_F$ flavor symmetry, irreducible $\mathrm{SU}(3)$, or pole models, which often provide differing predictions for these specific channels.

This analysis was performed using the combined datasets from Belle ($988.4 \text{ fb}^{-1}$) and Belle II ($427.9 \text{ fb}^{-1}$).
The signal decays were reconstructed using the following intermediate states: $\Lambda \to p \pi^-,\pi^0/\eta\to\gamma\gamma,\eta\to \pi^+\pi^-\pi^0$ and $\eta^{\prime}$ via both $\pi^+\pi^-\eta$ and $\pi^+\pi^- \gamma$.
To extract the signal yields, unbinned extended maximum-likelihood fits were performed on the invariant mass distributions of the $\Xi_c^0$ candidates as shown in Fig.~\ref{fig:xic_fits_scs}.
The $\Xi_c^0\to\Xi^- \pi^+$ decay was used as the normalization mode to determine the absolute branching fractions.

Significant signals were observed for the $\Lambda\eta$ mode with a statistical significance of $5.3\sigma$, while significances of $3.3\sigma$ and $1.4\sigma$ were obtained for the $\Lambda \eta^{\prime}$ and $\Lambda \pi^0$ modes, respectively.
Experimental systematic uncertainties are dominated by $\pi^0/\gamma$ reconstruction efficiencies and fitting model dependencies.
However, the external uncertainty of the normalization decay $\mathcal{B}(\Xi_c^0 \to \Xi^- \pi^+)$ constitutes the largest single contribution to the total systematic error.

This measurement constitutes the first evidence for the $\Xi_c^0 \to \Lambda \eta^{\prime}$ decay. As no significant signal was found for the $\Lambda\pi^0$ mode, an upper limit was established at the 90\% confidence level. These results, summarized in Table~\ref{tab:absolute_branching_ratios}, could provide new constraints on the relative contributions of factorizable and nonfactorizable diagrams in charmed baryon decays.

\begin{table}[h!tbp]
\centering
\caption{Absolute branching fractions for $\Xi_c^0$ decays. The uncertainties are statistical, systematic, and from $\mathcal{B}(\Xi_c^0 \to \Xi^-\pi^+)$, respectively.}
\label{tab:absolute_branching_ratios}
\begin{tabular}{ll}
\toprule
Mode & Branching Fraction \\
\midrule
$\mathcal{B}(\Xi_c^0 \to \Lambda\eta)$ & $(5.95 \pm 1.30 \pm 0.32 \pm 1.13) \times 10^{-4}$ \\
$\mathcal{B}(\Xi_c^0 \to \Lambda\eta^{\prime})$ & $(3.55 \pm 1.17 \pm 0.17 \pm 0.68) \times 10^{-4}$ \\
$\mathcal{B}(\Xi_c^0 \to \Lambda\pi^0)$ & $< 5.2 \times 10^{-4}$ (90\% C.L. limit) \\
\bottomrule
\end{tabular}
\end{table}

\subsection{$\Xi_c^+ \to pK_S^0, \Lambda \pi^+, \Sigma^0\pi^+$}
The study of SCS $\Xi_c^+$ decays into a light baryon and a meson provides essential data to test $\mathrm{SU}(3)_F$ flavor symmetry and various topological diagrammatic approaches in the charmed baryon sector.
These signal decays are SCS processes, where the interplay between various internal $W$-emission and $W$-exchange diagrams is expected to lead to non-zero branching fractions.

This analysis~\cite{Belle:2024xcs} utilized data samples from both Belle ($983.0 \text{ fb}^{-1}$) and Belle II ($427.9 \text{ fb}^{-1}$).
The signal decays were reconstructed via the following decay chains $K_S^0\to\pi^+\pi^-,\Lambda \to p \pi^-,$ and $\Sigma^0\to \Lambda \gamma$.
To determine absolute branching fractions, the yields were normalized to the normalization mode $\Xi_c^+ \to \Xi^- \pi^+ \pi^+$~\cite{Belle:2019bgi}.

We report the first observation of these three decay modes.
The statistical significances of all decay channels exceed $10\sigma$ in both Belle and Belle II data, with the exception of the $\Xi_c^+ \to \Lambda \pi^+$ mode in the Belle dataset, which has a significance of $7.6\sigma$.
For the $pK_S^0$ and $\Lambda \pi^+$ mode, the dominant systematic uncertainties originate from the fitting method and Dalitz efficiency corrections.
In the $\Sigma^0\pi^+$ mode, the fitting uncertainty is the leading source, followed by comparable contributions from $\gamma$ reconstruction and Dalitz efficiency corrections.

These results show varying degrees of agreement and tension with theoretical predictions, including $\mathrm{SU}(3)_F$ flavor symmetry models~\cite{Zhong:2022exp, Hsiao:2021nsc} and topological diagrammatic or irreducible $\mathrm{SU}(3)$ approaches~\cite{Zhong:2024qqs}.
These measurements provide critical constraints for $\mathrm{SU}(3)$ amplitudes in the charmed baryon system.

\begin{table}[h!tbp]
\centering
\caption{Absolute branching fractions for $\Xi_c^+$ decays. The uncertainties are statistical, systematic, and external from $\mathcal{B}(\Xi_c^+ \to \Xi^-\pi^+\pi^+)$.}
\label{tab:xicp_absolute}
\begin{tabular}{lc}
\toprule
Mode & Branching Fraction $[10^{-4}]$ \\
\midrule
$\Xi_c^+ \to pK_S^0$      & $7.16 \pm 0.46 \pm 0.20 \pm 3.21$ \\
$\Xi_c^+ \to \Lambda\pi^+$ & $4.52 \pm 0.41 \pm 0.26 \pm 2.03$ \\
$\Xi_c^+ \to \Sigma^0\pi^+$ & $1.20 \pm 0.08 \pm 0.07 \pm 0.54$ \\
\bottomrule
\end{tabular}
\end{table}

\subsection{$\Xi_c^+ \to \Sigma^+K_S^0, \Xi^0\pi^+, \Xi^0K^+$}

This study~\cite{Belle:2025skz} presents improved measurements of the branching fractions for the CF decays $\Xi_c^+ \to \Sigma^+ K_S^0$ and $\Xi_c^+ \to \Xi^0\pi^+$. Additionally, the first observation of the SCS decay $\Xi_c^+\to \Xi^0 K^+$ is reported.

The study utilizes data samples collected by the Belle ($983.0 \text{ fb}^{-1}$) and Belle II ($427.9 \text{ fb}^{-1}$) experiments.
To determine the branching fraction ratios, the decay $\Xi_c^+ \to \Xi^-\pi^+\pi^+$ is employed as the normalization mode.
The SCS decay $\Xi_c^+ \to \Xi^0 K^+$ is observed for the first time with statistical significances of $4.7\sigma$ in Belle data and $5.5\sigma$ in Belle II data.
The fitted signal yields for Belle and Belle II are, respectively, ($288 \pm 41$ and $182 \pm 31$) for $\Sigma^+ K_S^0$, ($2782 \pm 74$ and $1469 \pm 40$) for $\Xi^0 \pi^+$, and ($138 \pm 31$ and $100 \pm 20$) for $\Xi^0 K^+$.

The measured absolute branching fractions are summarized in Table~\ref{tab:xicp_normalized}.
The systematic uncertainties for the three decay modes are dominated by the signal extraction method and the $\pi^0$ reconstruction efficiency.
The results are broadly consistent with several theoretical predictions, specifically models utilizing $\mathrm{SU}(3)_F$ symmetry with an irreducible approach~\cite{Geng:2019xbo,Huang:2021aqu}.

\begin{table}[h!tbp]
\centering
\caption{Branching fractions for $\Xi_c^+$ decays normalized to the reference mode $\mathcal{B}(\Xi_c^+ \to \Xi^-\pi^+\pi^+)$. The uncertainties are statistical, systematic, and from the reference mode.}
\label{tab:xicp_normalized}
\begin{tabular}{lc}
\toprule
Mode & Absolute Branching Fraction [\%] \\
\midrule
$\Xi_c^+ \to \Sigma^+ K_S^0$ & $0.194 \pm 0.021 \pm 0.009 \pm 0.087$ \\
$\Xi_c^+ \to \Xi^0 \pi^+$    & $0.728 \pm 0.014 \pm 0.027 \pm 0.326$ \\
$\Xi_c^+ \to \Xi^0 K^+$      & $0.049 \pm 0.007 \pm 0.003 \pm 0.022$ \\
\bottomrule
\end{tabular}
\end{table}

\subsection{$\Lambda_c^+ \to p K_S^0 \pi^0$}
Using $980 \text{ fb}^{-1}$ of Belle data, the branching fraction of $\Lambda_c^+ \to p K_S^0 \pi^0$ relative to $\Lambda_c^+ \to p K^- \pi^+$ was measured.
This study~\cite{Belle:2025voy} directly examines the isospin properties of the weak interaction and significantly improves upon the precision of previous measurements.

The signal yield is extracted relative to the well known normalization mode $\Lambda_c^+ \to p K^- \pi^+$.
The measured ratio of branching fractions is determined to be $0.339\pm0.002 \pm 0.009$, where the uncertainties are statistical and systematic, respectively.
The systematic uncertainties are primarily driven by the reconstruction efficiencies of the $K_S^0$ and $\pi^0$ candidates, contributing approximately 1.5--1.6\% for each component to the relative branching fraction.
Using the world average value~\cite{ParticleDataGroup:2024cfk} for the normalization mode, $\mathcal{B}(\Lambda_c^+ \to pK^-\pi^+) = (6.24 \pm 0.28)\%$, the absolute branching fraction is calculated to be $(2.12 \pm 0.01 \pm 0.05 \pm 0.10)\%$, where the third uncertainty accounts for the normalization mode.

Furthermore, this study investigated the resonance structures within the Dalitz phase space.
It reveals a distinct peaking structure in $p\pi^0$ invariant mass spectrum near the $p \eta$ threshold.
The structure potentially indicates a threshold cusp effect associated with the $N(1535)^+$ resonance.
Such findings emphasize the necessity for a comprehensive amplitude analysis in future studies to fully understand the underlying decay dynamics.

\section{Search for CP violation}
We report the first individual $A_{CP}$ measurements for three-body charmed baryon decays using $428~\text{fb}^{-1}$ of Belle~II data~\cite{Belle-II:2025xvc}. 
The signal decays consist of $\Xi_c^+\to\Sigma^+h^+h^-$ and $\Lambda_c^+ \to p h^+ h^-$, where $h$ denotes a $\pi$ or $K$ meson.

The raw asymmetry, $A_{N}(X_c^+)$, is defined by the signal yields $N$ of the decay and its charge conjugate.
Production asymmetries are eliminated by averaging measurements from the forward and backward center-of-mass hemispheres, $A_{N}^{\prime}$.
Instrumental asymmetries are handled by subtracting control channel yield asymmetries:
\begin{equation}
\begin{split}
    A_{CP}(\Xi_c^+\to \Sigma^+ h^+ h^-)=A^{'}_N(\Xi_c^+ \to \Sigma^+ h^+ h^-) \\
    - A^{'}_N(\Lambda_c^+ \to \Sigma^+ h^+ h^-) \\
\end{split}
\end{equation}
\begin{equation}
\begin{split}
    A_{CP}(\Lambda_c^+ \to p h^+ h^-)
    &=A^{'}_N(\Lambda_c^+\to p h^+ h^-) \\
    & -A^{'}_N(\Lambda_c \to p \pi^+ K^-) \\
    & -A^{'}_N(D^0 \to \pi^+K^- \pi^+ \pi^-)
\end{split}
\end{equation}

All measured asymmetries are consistent with $CP$ conservation, shown in Table~\ref{tab:cp_asymmetry}.
Systematic uncertainties primarily arise from fitting method and $h^+h^-$ detection asymmetries, which range from 0.2\% to 0.4\%.
Using these measurements to test the $U$-spin symmetry sum rules, we obtain $A_{CP}(\Xi_c^+ \to \Sigma^+ \pi^+ \pi^-) + A_{CP}(\Lambda_c^+ \to p K^+ K^-) = (13.4 \pm 7.0 \pm 0.9)\%$ and $A_{CP}(\Xi_c^+ \to \Sigma^+ K^+ K^-) + A_{CP}(\Lambda_c^+ \to p \pi^+ \pi^-) = (4.0 \pm 6.6 \pm 0.7)\%$.
Both results are consistent with $U$-spin symmetry within the current statistical precision.

\begin{table}[h!tbp]
\centering
\caption{Measured $CP$ asymmetries ($A_{CP}$). Uncertainties are statistical and systematic.}
\label{tab:cp_asymmetry}
\begin{tabular}{lc}
\toprule
Mode & $A_{CP}$ [\%] \\
\midrule
$\Xi_c^+ \to \Sigma^+ K^- K^+$ & $3.7 \pm 6.6 \pm 0.6$ \\
$\Xi_c^+ \to \Sigma^+ \pi^- \pi^+$ & $9.5 \pm 6.8 \pm 0.5$ \\
$\Lambda_c^+ \to p K^- K^+$ & $3.9 \pm 1.7 \pm 0.7$ \\
$\Lambda_c^+ \to p \pi^- \pi^+$ & $0.3 \pm 1.0 \pm 0.2$ \\
\bottomrule
\end{tabular}
\end{table}

\section{Summary}
Utilizing Belle and Belle II (Run 1) data, we have presented first observations and improved measurements of several $\Xi_c^{0/+}$ and $\Lambda_c^+$ decay modes.
These results provide vital constraints for $\mathrm{SU}(3)_F$, nonfactorizable and other models.
Searches for CP violation in three-body charm baryon decays show no evidence of asymmetry, consistent with $U$-spin symmetry predictions.
As Belle II Run 2 data collection continues, the targeted $1~\text{ab}^{-1}$ dataset in 2026 will enable even more stringent tests of $CP$ violation and the observation of rare decay modes that are currently statistically limited.




\nocite{*}
\bibliographystyle{elsarticle-num}
\bibliography{jos}







\end{document}